\documentclass[conference,compsoc]{IEEEtran}

\usepackage{amsmath}
\usepackage{graphicx}
\usepackage{subcaption}
\usepackage{color}
\usepackage{hyperref}

\title{An Ensemble Framework for Detecting Community Changes in Dynamic Networks}
\author{\IEEEauthorblockN{Timothy La Fond, Geoffrey Sanders, Christine Klymko, and Van Emden Henson}
\IEEEauthorblockA{Lawrence Livermore National Laboratory\\
Livermore, California 94550\\
}}

\begin{document}
\maketitle

\begin{abstract}
Dynamic networks, especially those representing social networks, undergo constant evolution of their community structure over time.  Nodes can migrate between different communities, communities can split into multiple new communities, communities can merge together, etc.  In order to represent dynamic networks with evolving communities it is essential to use a dynamic model rather than a static one.  Here we use a dynamic stochastic block model where the underlying block model is different at different times.  
In order to represent the structural changes expressed by this dynamic model the network will be split into discrete time segments and a clustering algorithm will assign block memberships for each segment.  In this paper we show that using an ensemble of clustering assignments accommodates for the variance in scalable clustering algorithms and produces superior results in terms of pairwise-precision and pairwise-recall.  We also demonstrate that the dynamic clustering produced by the ensemble can be visualized as a flowchart which encapsulates the community evolution succinctly.
\end{abstract}

\section{Introduction}

%Network analysis, clustering is useful
%
%Real networks are dynamic and have changing underlying model
%
%Citations other dynamic models?  GraphScope, other evolution papers
%
%Need approaches that find transition point of the cluster structure
%
%Visualization of community evolution useful too
%
%

Over the past decade or so, network analysis has emerged as an important area of mathematics and computer science, as well as in many other disciplines (see \cite{boccaletti, brandes, Newman} among many others).  One of the significant questions in network analysis is how to cluster the nodes in the network to identify underlying communities \cite{fortunato, malliaros}.  Clustering static graphs reveals important community structure that can be leveraged for many important analysis tasks involving relational datasets.  
Several fast clustering techniques have been defined for detecting communities that have high internal connectivity relative to their external connectivity \cite{blondel, peixoto1,peixoto2,peixoto3}.
Analysis of community structure in {\em dynamic graphs}, where each edge is associated with a timestamp, is vastly more complex.
Many real-world networks experience concept drift as the networks evolve; such evolutions cannot be accurately represented with a static model. Ignoring the temporal information and applying static clustering techniques entire dataset risks missing important structural changes; important but transient community development could remain undiscovered.

\begin{figure}[h]
\begin{center}
\includegraphics[width=0.4\textwidth]{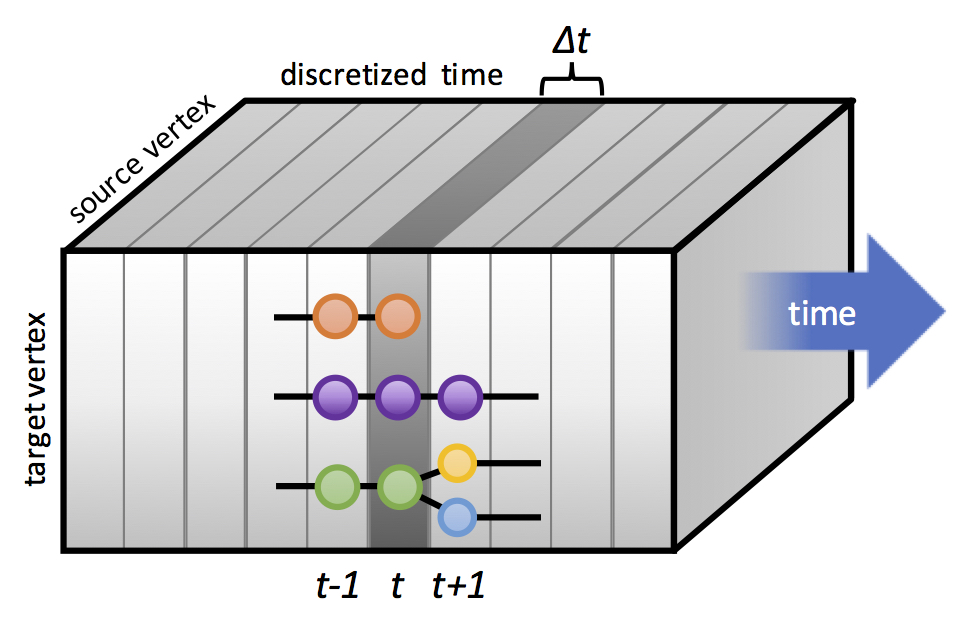}
\end{center}
\caption{\small Schematic of the dynamic community detection framework employed. For each timestep, $t$, a representative partitioning is determined.   Representative partitionings from adjacent timesteps $(t-1)$ and $t+1$ are compared for tracking communities and their dynamic structural changes such as community death (dark orange) or community splitting (green into blue and yellow). }
\label{fig:framework}
\end{figure}

Several types of changes in community structure are of high analytic interest, including (i) the birth/death of a community \cite{greene}, (ii) the growth/decay of a community \cite{greene}, (iii) the changes of internal connectivity of a community (such as a dense community becoming increasingly bipartite \cite{sun}), (iv) groups of communities splitting/merging \cite{greene, hong}, and (v)  member migration between groups of two or more communities (as well as the identification of nodes that do not migrate \cite{pandit}). %\cite{chakraborty}).
Clearly, there is a need for both the ability to perform {\em dynamic community detection} to find the community behavior at any given time and to {\em represent community evolution} to display the kinds of temporal effects that occur in the network.
In this paper, we design a framework for taking a static community detection algorithm and employing it to find the evolving community structure of dynamic networks.  Our framework also provides a interpretable visualization of the major changes to the community structure.

%Here, we discuss the design of a framework for detecting transient communities and community dynamics, where retention of the temporal information is necessary to detect the time-dependent phenomenon.

Some dynamic community detection algorithms, for example, GraphScope, have memory and will continue to use the discovered data model from prior time steps as long as that model passes some performance metrics \cite{sun}.  This memory effect can sometimes be detrimental if the algorithm fails to update the model during a change in the underlying structure.  Other algorithms have no memory at all and merely connect the results of independent community detection passes run at different times in the network \cite{greene}.  We will describe how to apply our framework in both a memory and memoryless fashion, where the version with memory attempts to smooth the temporal noise of the data at the cost of sometimes lagging behind immediate changes to the underlying model.

Often edge timestamps are {\em essentially continuous}, where the resolution of timestamps is on much smaller intervals than the span of time required to resolve communities, both statically and during their evolution.
For example, we might have timestamps that have resolution in the seconds while a few hours of connectivity data are necessary to resolve most community structure present in the dataset.
A large class of dynamic graph analysis techniques discretize the time dimension into a number of uniform non-overlapping {\em time segments}.

%Underlying communities may involve different densities due to the participation rates of their members and may not be visible at the resolution provided by the timestep width appropriate for the rest of the graph.
%Underlying communities may involve different densities, due to the participation rates of their members, and the timestep size required to detect various communities may be staggeringly different.   % add example?
%The requirements on the timestep size to see different community dynamics (i)-(v) in hetergeneous data are even more restrictive, and in general it is likely impossible to detect all these events in a large dynamic graph with a single timestep size. 
%Truly practical dynamic community anaylsis tools must also discover the relevant timestep sizes.
%As this issue is an open problem, in this work we assume that we are given an appropriate timestep width and focus on exploring the community structure changes that are detectable in this temporal discretization.  
The selection of an appropriate timestep width which provides sufficient resolution to detect temporal events remains an open problem, with time segments typically being defined to start and end at a natural frequency determined by the data source, e.g. daily or weekly.
%We will be adhering to this rule of thumb and using weekly time segments for the real world data we use in Sections \ref{sec:semireal_experiments} and \ref{sec:real_experiments}.
We will also restrict ourselves to community dynamics (i) and (iv) (birth, death, splitting, and merging).
%In this work, we assume we are given a timestep size, and we aim to expose community structural changes that are detectable in this temporal discretization.
%[TODO: decide if we want to talk about this issue (graph resolution) at all: something of a tangent, and not a focus of the current paper, we assume the timestamp width is appropriate]

In this paper we:

\begin{itemize}
\item Define a dynamic stochastic block model capable of producing data with community evolution events.
\item Introduce a {\em Dynamic Clustering Ensemble Framework} which, given a baseline static clustering algorithm, finds the dynamic community behavior of a network and reports any evolution in the community structure.
\item Demonstrate that the ensemble framework has superior pairwise-precision and pairwise-recall performance in recovering block assignments than than the baseline algorithm. We apply this ensemble to synthetic data, drawn from both static and dynamic models, as well as to semi-synthetic data.
\item Create a visualization of the community evolution in a real world dataset and illustrate the types of community dynamics found (Figure \ref{experimentEU}).
\end{itemize}

\section{Approach}

\subsection{Data Model}

%Dynamic graph models are useful for algorithmic development: 
To aid in the development of temporal algorithms, generative dynamic graph models are often used.  These models support:
(a) building fully synthetic examples with various community dynamics, allowing algorithm validation on examples with strong ground-truth; 
(b) injecting synthetic community dynamics into real world dynamic graphs, allowing algorithm validation in the face of realistic noise; and
(c) gauging the statistical significance of various community events that have been detected.
Degree-corrected stochastic block models whose model parameters are functions of continuous time are able to exhibit the kinds of phenomenon described in (i)-(v) above, and we will be using a version for our synthetic and semi-synthetic data analysis.
%Note that the mixed-membership property is essential for (ii) and (v), as well as modeling community overlap.
%In \S2 we describe the degree-corrected model we use for the experimental validation. 

In order to create the type of community dynamics we expect to see in real networks we introduce a dynamic stochastic block model as a generative graph model where every edge $A_{ijt}$ is sampled according to a Poisson distribution with a time-dependent rate $\lambda_{ij}$:
\[ A_{ijt} \sim Poisson(\lambda_{ij}(t)) \\ 
\qquad \qquad
\lambda_{ij}(t) = \theta_i \theta_j \Sigma_{b_i,b_j}(t) \]
where $\theta_i$ controls node $i$'s expected degree, $b_i$ is $i$'s block assignment, and $\Sigma_{b_i,b_j}(t)$ is the interaction strength between blocks $b_i$ and $b_j$ at time $t$.  
%We will extend the capabilities of this model by replacing $\Sigma_{b_i,b_j}$ with a time-varying function $F(b_i,b_j,t)$ which allows for the block model interactions to change over time.  
As the block model is a function of time, by gradually changing the edge probability between collections of nodes over time we can introduce smooth splits and merges into the community structure.  
For example, if the value of $\Sigma_{b_i,b_j}(t)$ grows linearly with increasing $t$ we effectively cause $b_i$ and $b_j$ to merge into a single block as their interconnectivity grows.  

%We will use this capability to show the performance of the clustering algorithm using our framework when these community evolutions exist in the data.

\subsection{Ensemble Framework}

Given a dynamic graph as defined above and an \emph{a priori} division of the graph stream into discrete time segments, the algorithm presented clusters the graph at every segment, using as a baseline the Markov chain Monte Carlo (MCMC) algorithm described in the works of Peixoto \cite{peixoto1,peixoto2,peixoto3}.  This algorithm uses an ergodic Markov chain to randomly modify the block membership of each node and probabilistically accepts or rejects each modification.  Although the algorithm will converge given a sufficiently long mixing time, the required mixing times are often prohibitively long and, in practice, approximations to the MCMC process are used.  These approximations, combined with the non-determinism of the MCMC process, lead to variability in the produced clustering given different runs of the MCMC algorithm on the same data.

\begin{figure}[h]
\begin{center}
\includegraphics[width=0.3\textwidth]{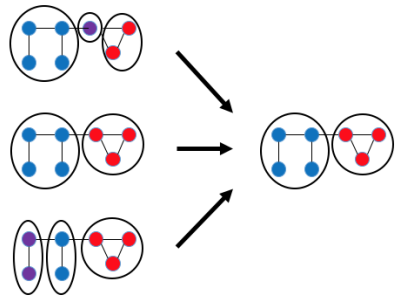}
\end{center}
\caption{\small The clusterings determined by various runs of the MCMC clustering algorithm are used to form a representative clustering of the nodes.}
\label{fig:representative}
\end{figure}

We address this variability using an ensemble approach which performs multiple runs of the MCMC clustering algorithm at each time segment.  This accommodates for variance in the clustering algorithm by applying the algorithm multiple times and obtaining a cloud of possible partitions.  These clouds are then resolved to produce a best overall cluster assignment for the time segment, referred to as the \emph{representative clustering}, which corresponds to the most typical partitions seen in the cloud.  A small example of this can be seen in Figure \ref{fig:representative}.

Let $H(X,Y)$ be any similarity function across two sets of nodes.  
%Given $I$ runs of the MCMC clustering algorithm at time $t$, leading to $I$ different partitions of the form $\{B^{i,t}_{1}, B^{i,t}_{2}... B^{i,t}_{c_i},  \}$ for $i = 1$ to $I$, the algorithm reports the partition $B^t = \{ B^t_1, B^t_2... B^t_{\tilde{c} } \}$ that satisfies 
Given $K$ runs of the MCMC clustering algorithm at time $t$, leading to $K$ different partitions of the form $\{B^{k,t}_{1}, B^{k,t}_{2}... B^{k,t}_{c_k},  \}$ for $k = 1$ to $K$, the algorithm reports the partition $B^t = \{ B^t_1, B^t_2... B^t_{\tilde{c} } \}$ that satisfies 

\[ B^t = \left\{ \max_{B^{k,t}_x} \mathbf{E}_{l \neq k}\left[ \max_{y} H(B^{k,t}_x, B^{l,t}_y)  \right] \right\} \] 
where $B^{k,t}_x$ is the set of nodes in the $x$th block of the $k$th partition at time $t$, $c_k$ is the number of blocks in partition $k$, $\tilde{c}$ is the median number of blocks across all $K$ partitions of time $t$, and ${\bf E}_{k \neq l}[\cdot]$ is the expectation taken over all pairs of different iterations. 
Here we use similarity function $H(X,Y) =\frac{| X \cap Y |}{| X \cup Y |}$. %\ctxt{[TODO: what do we do with $H(X,Y)$ now that we have defined it?]}  

%<<<<<<< HEAD
%This resolution step can also be extended to ``smooth" the output by incorporating $B^{j,t-1}_y$ and/or $B^{j,t+1}$ into the bounds of the expectation, emphasizing connections between clusters over time if an algorithm with memory is desired.  This smoothing approach compares the possible partitions available for time step $t$ to those from $t-1$ and $t+1$ in order to find blocks which overlap significantly across different time steps.  %allowing $B^j_y$ to come from a partition $j$ which was generated at a different time than partition $i$.  
%=======
This resolution step can also be extended to ``smooth" the output by incorporating $B^{l,t-1}_y$ and/or $B^{l,t+1}_y$ (as well as additional time segments in each direction) into the bounds of the expectation, emphasizing connections between clusters over time if an algorithm with memory is desired.  This smoothing approach compares the possible partitions available for time step $t$ to those from $t-1$ and $t+1$ in order to find blocks which are consistent across different time steps.  %allowing $B^j_y$ to come from a partition $j$ which was generated at a different time than partition $i$.  
%>>>>>>> 19faaad50817cf560aa4919508b782e8f20a18aa

A slight overhead is incurred during the resolution of the ensemble of partitions. For a given time step, each partition $k$ out of the $K$ different partitions of the graph has $c_k$ blocks, and all pairs of blocks from different partitions must be compared for the intersection of their member nodes.  If using a hashed implementation, each intersection operation costs $\Theta(min(n_x, n_y))$, where $n_x$ and $n_y$ are the size of blocks $x$ and $y$.  For every pair of partitions, all pairs of block must be compared, and each are an average size of $\frac{N}{c_k}$ or $\frac{N}{c_l}$.  The cost of all comparisons is then 

\[ \Theta\left( c_k c_l K (K-1) min\left(\frac{N}{c_k}, \frac{N}{c_l}\right)\right) = \Theta\left( N \cdot min(c_k, c_l) \right) .\]
Assuming that the number of blocks returned in any partition is less than $log^2 N$, the computational complexity of the ensemble resolution step is equal to or less than the cost of clustering the graph once at any particular time segment on average.  
Each iteration of the graph clustering algorithm in the ensemble can be run in parallel, needing no communication across the iterations until the resolution step, making the cost of each time segment $\Theta\left( N log^2 N \right)$.
In the memoryless, unsmoothed version of the framework the individual time segments are also independent and can be run in parallel.  The smoothed version requires communication between adjacent time segments after graph clustering but before the cluster resolution step, where the clouds of clusterings obtained by the $K$ iterations are exchanged.

%\ctxt{Since running the MCMC ensemble is trivially parallelizable, the overall computational complexity is not increased from that of the base algorithm. [TODO: this previous setence needs work.   For one, combining the ensemble is not embarassing parallel.]}  %Assuming that the number of blocks found in any of the partitions is has a sublinear relationship to 

\subsection{Visualization of Output}

Finally, after producing a representative clustering at each time segment with smoothing between adjacent segments, we produce a visualization showing how the community structure changes over time.  
An example of the visualization obtained from a real world network can be seen in Figure \ref{experimentEU}.
In this community graph, the nodes represent communities with the node diameter indicating community size, and the communities are connected in sequential time segments if they share members with a thicker edge indicating more overlap between the two sets of members.  This visualization allows for a greater understanding of the evolution of community structure than merely reporting on the number of communities at each time segment.  Through it, community evolution can be observed, including events such as communities merging, splitting, forming out of background noise, and dissolving into background noise.

%Slight overhead in resolving cluster ensemble: $\overline{B} \#B K$, where $\overline{B}$ is average block size, $\#B$ is average number of blocks per iteration, 
 
\section{Experiments}

To validate our method we performed a series of experiments on synthetic, semi-synthetic, and real-world datasets.  We evaluated the clustering methods using both pairwise-precision and pairwise-recall, as well as number of blocks reported where applicable.  The pairwise-precision measures the percentage of node-pairs which are correctly identified as being in the same cluster: (true positives)/(true positives + false positives).  Pairwise-recall considers all pairs of nodes belonging to the same ground-truth cluster and measures the fraction of them which are in the same identified cluster: (true positives)/(true positives + false negatives).  More information can be found in \cite{rand, hubert}.

\subsection{Synthetic Experiments}

\begin{figure*}
\begin{subfigure}[b]{0.33\textwidth} 
\includegraphics[width=\textwidth]{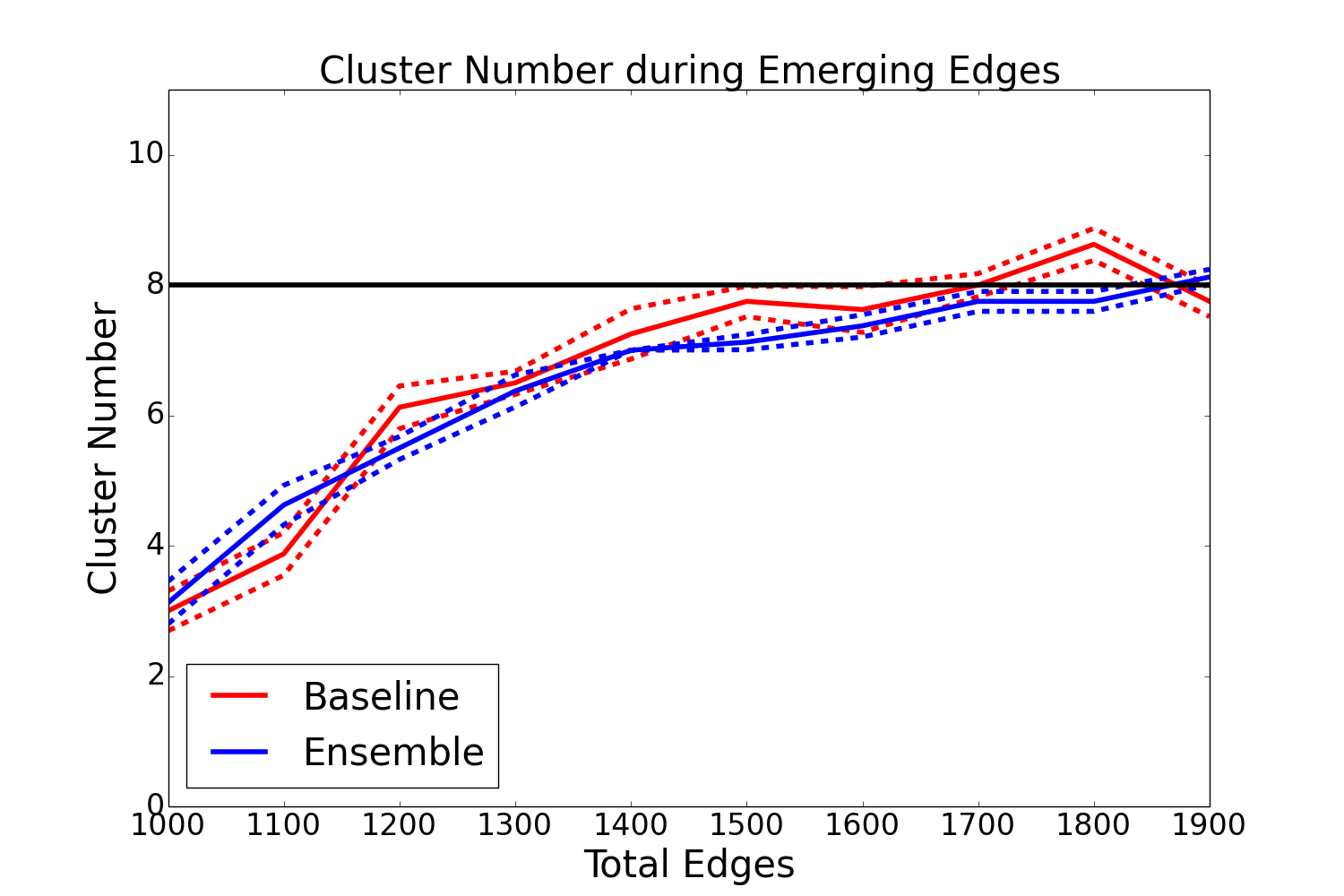}
\end{subfigure}
\begin{subfigure}[b]{0.33\textwidth} 
\includegraphics[width=\textwidth]{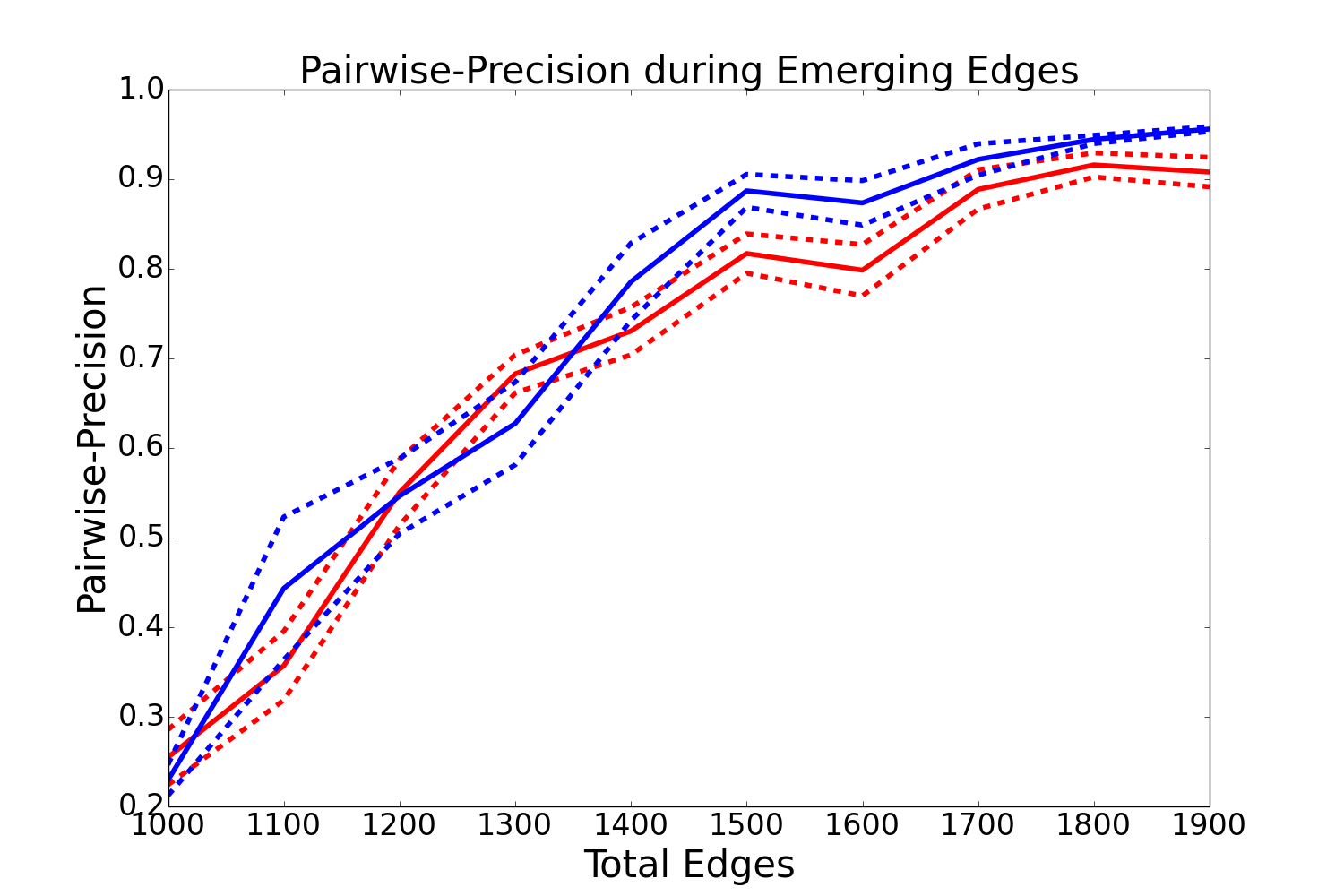}
\end{subfigure}
\begin{subfigure}[b]{0.33\textwidth} 
\includegraphics[width=\textwidth]{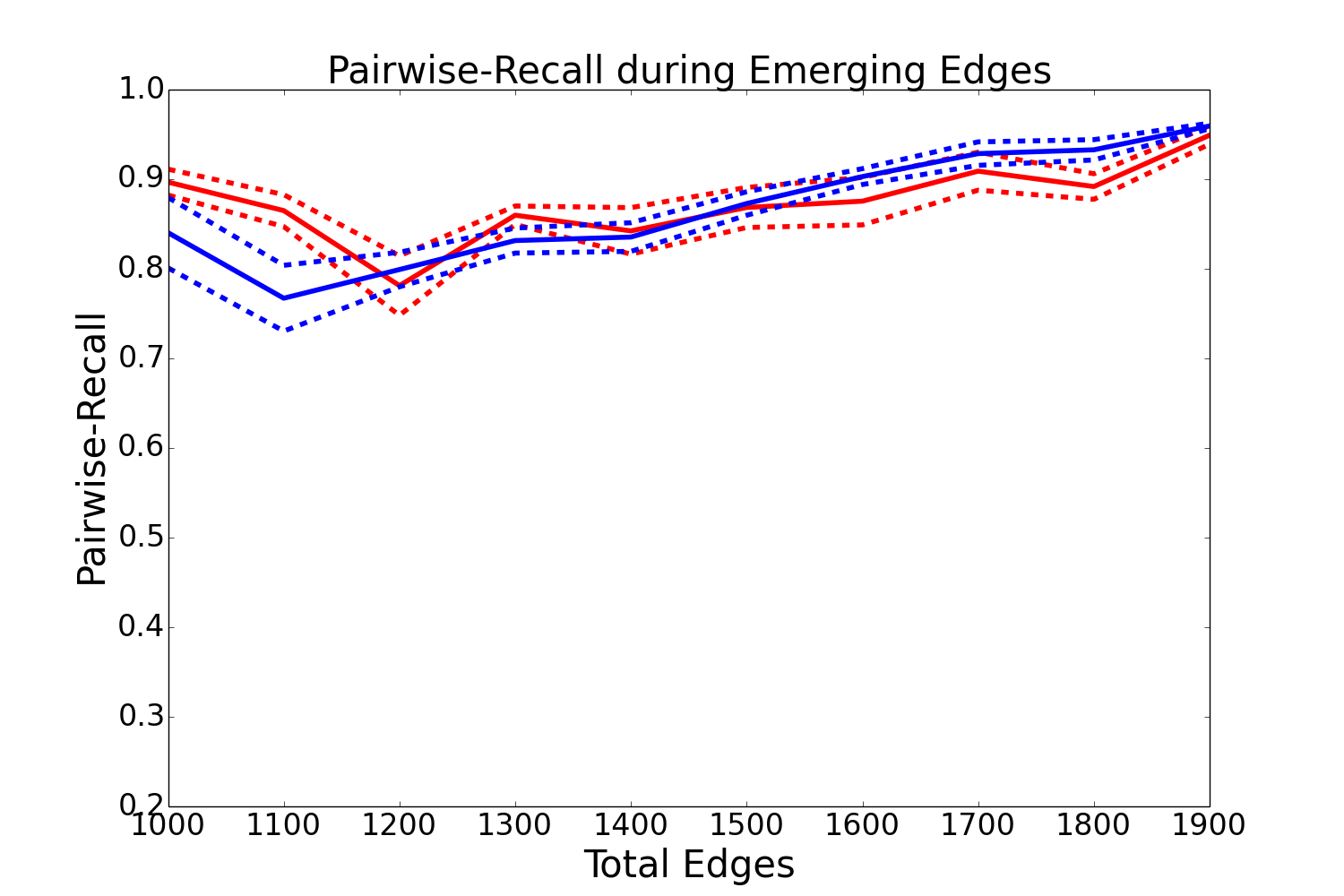}
\end{subfigure}
\caption{\small Performance on emerging edges experiment.  Red lines represent the baseline algorithm and blue represent the ensemble approach, with dashed lines showing the standard error over 10 runs of the algorithms.  The left-hand panel shows the number of blocks reported (compared to ground truth shown by the black line) as more edges are sampled.  Pairwise-precision among nodes as more edges are sampled is shown in the middle panel.  The right-hand panel shows the change in pairwise-recall among nodes as more edges are sampled.}
\label{experimentemerging}
\end{figure*}

% \ctxt{[TODO: font size on figures needs to be bigger; hard to see when printed; same for next two figures.]} 

The first experiment comparing the ensemble model to the baseline is performed on a synthetic dataset where edges emerge from a network stream over time and the clustering uses all edges observed up to the current time.  The block model from which the edges are sampled is considered constant; observing more edges merely increases the sample size and considerations like model drift are ignored.  The data consists of 500 node synthetic graphs: each of these has 8 ground truth blocks with a 1:5 external/internal edge ratio, and the number of edges observed was scaled from 1000 to 1900. 

The results from this experiment are shown in Figure \ref{experimentemerging}.  Red lines represent the baseline MCMC algorithm while blue lines are the ensemble algorithm; smoothing was not used as it is designed for dynamic models not static ones.  The dashed lines show the standard errors over ten runs of the algorithms.  The left-hand panel reports the number of clusters found by both algorithms.  Here, the black line shows the true number of clusters.  The middle panel and right-hand panels report the pairwise-precision and pairwise-recall, respectively, as more edges are sampled.  With fewer than approximately 1400 edges sampled neither method has good performance due to the number of blocks being under estimated.  However, as more edges are sampled, the ensemble method has strictly superior pairwise-precision.  The pairwise-recall performance appears about the same for both the baseline and the ensemble.

\begin{figure}[h]
\begin{center}
\includegraphics[width=0.4\textwidth]{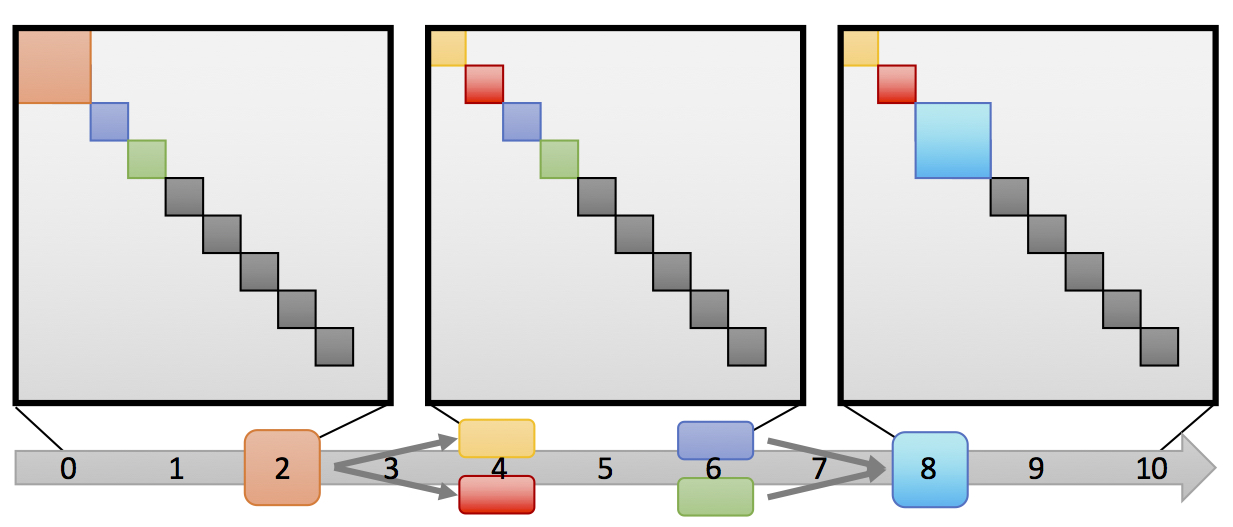}
\end{center}
\caption{\small A high level view of the evolving community dynamics from the second experiment on synthetic data.  The blocks are shown in the ``spy plot'' of the adjacency matrices associated with the underlying block model.}
\label{fig:synth2}
\end{figure}

The second experiment is another synthetic dataset, but here the edges are drawn from a dynamic model where the block structure is changing over time.  The model consists of a 100-node block which splits into two 50-node blocks between time segments 2 and 4; two 50-node blocks which merge into one 100-node block between time segments 6 and 8; five 50-node blocks which remain constant; and 50 nodes which connect to any node with equal probability, for a total of 500 nodes.  2000 edges were sampled for each time segment.  A high level view of these dynamics is shown in Figure \ref{fig:synth2}. 

\begin{figure*}
\begin{subfigure}[b]{0.33\textwidth} 
\includegraphics[width=\textwidth]{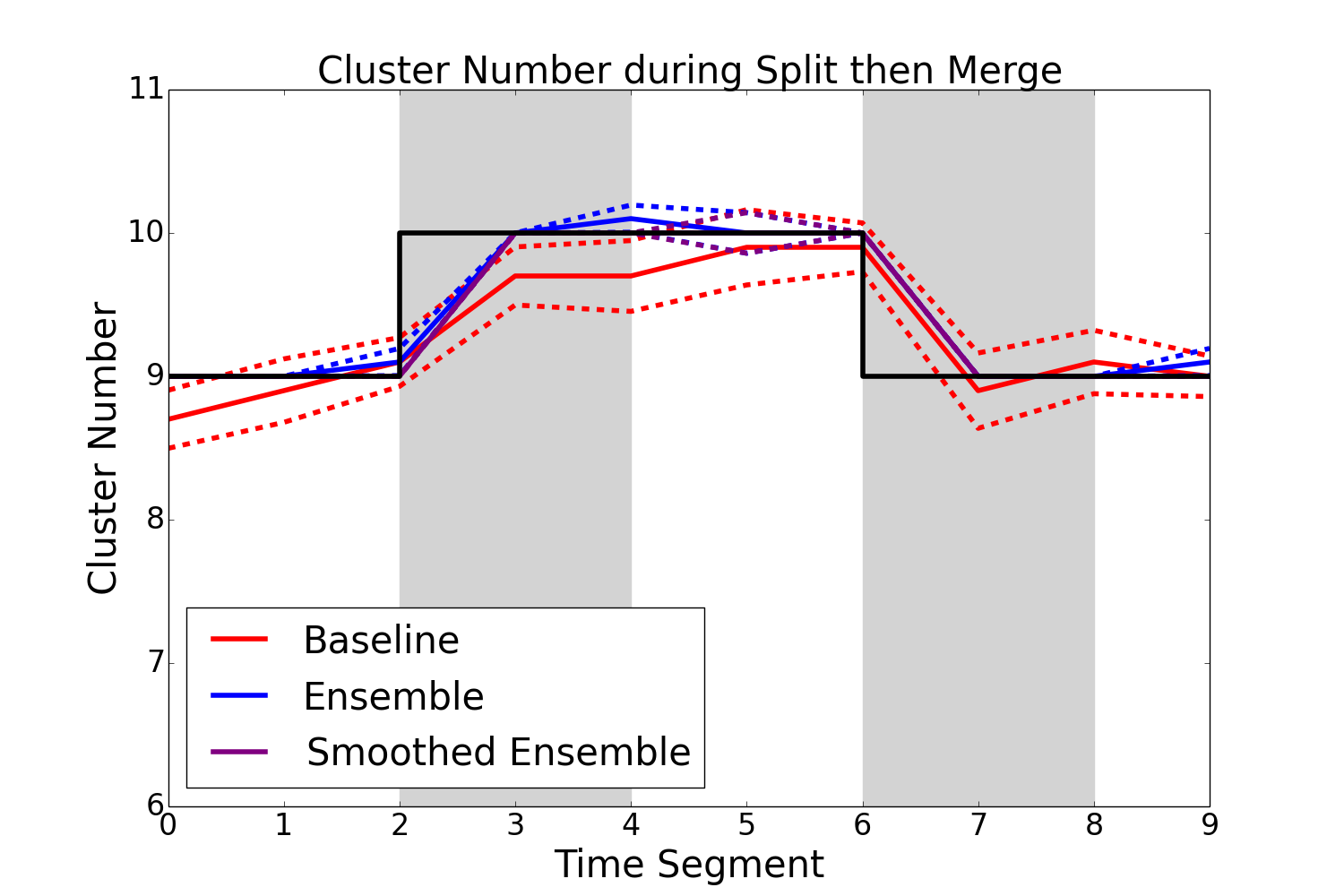}
\end{subfigure}
\begin{subfigure}[b]{0.33\textwidth} 
\includegraphics[width=\textwidth]{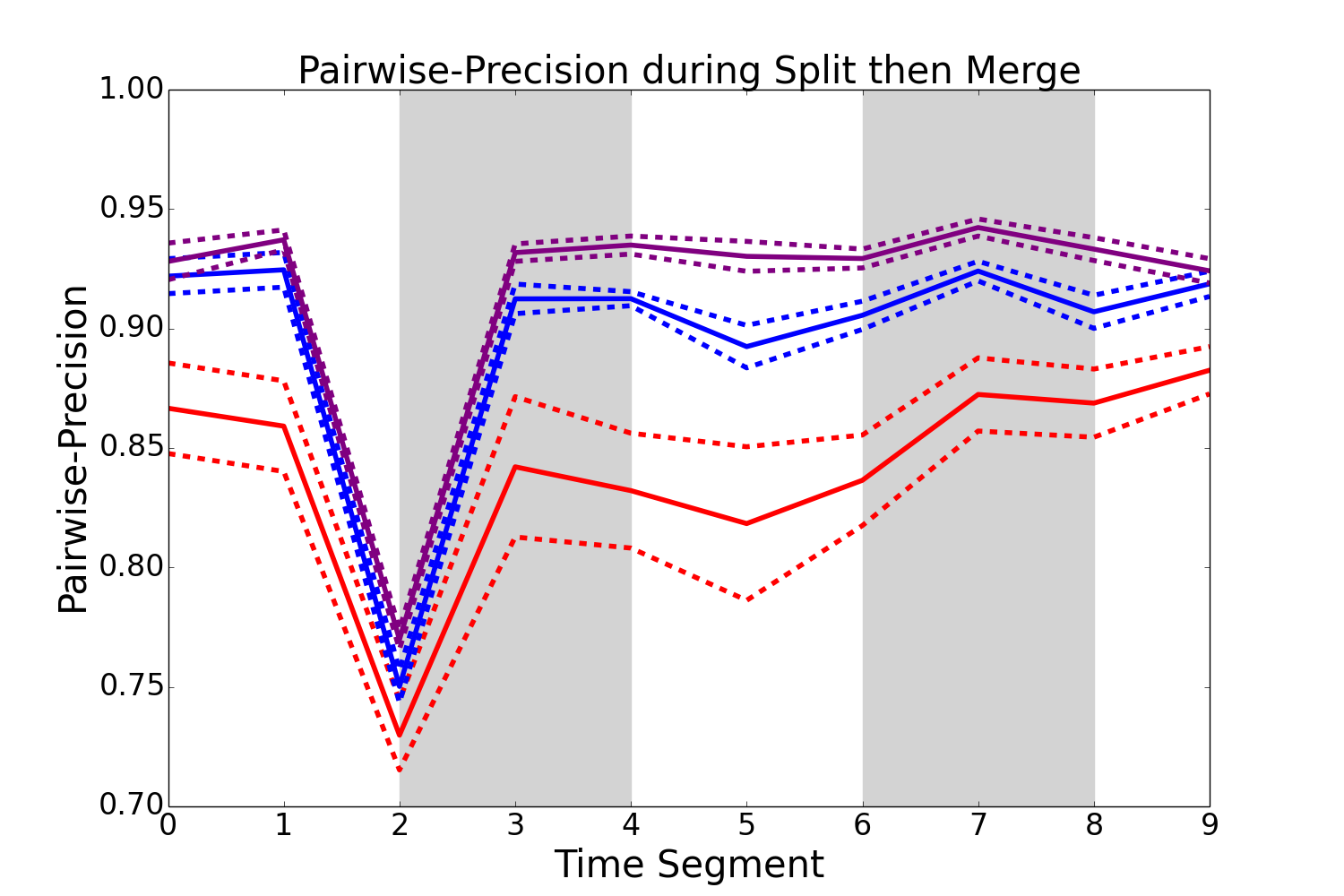}
\end{subfigure}
\begin{subfigure}[b]{0.33\textwidth} 
\includegraphics[width=\textwidth]{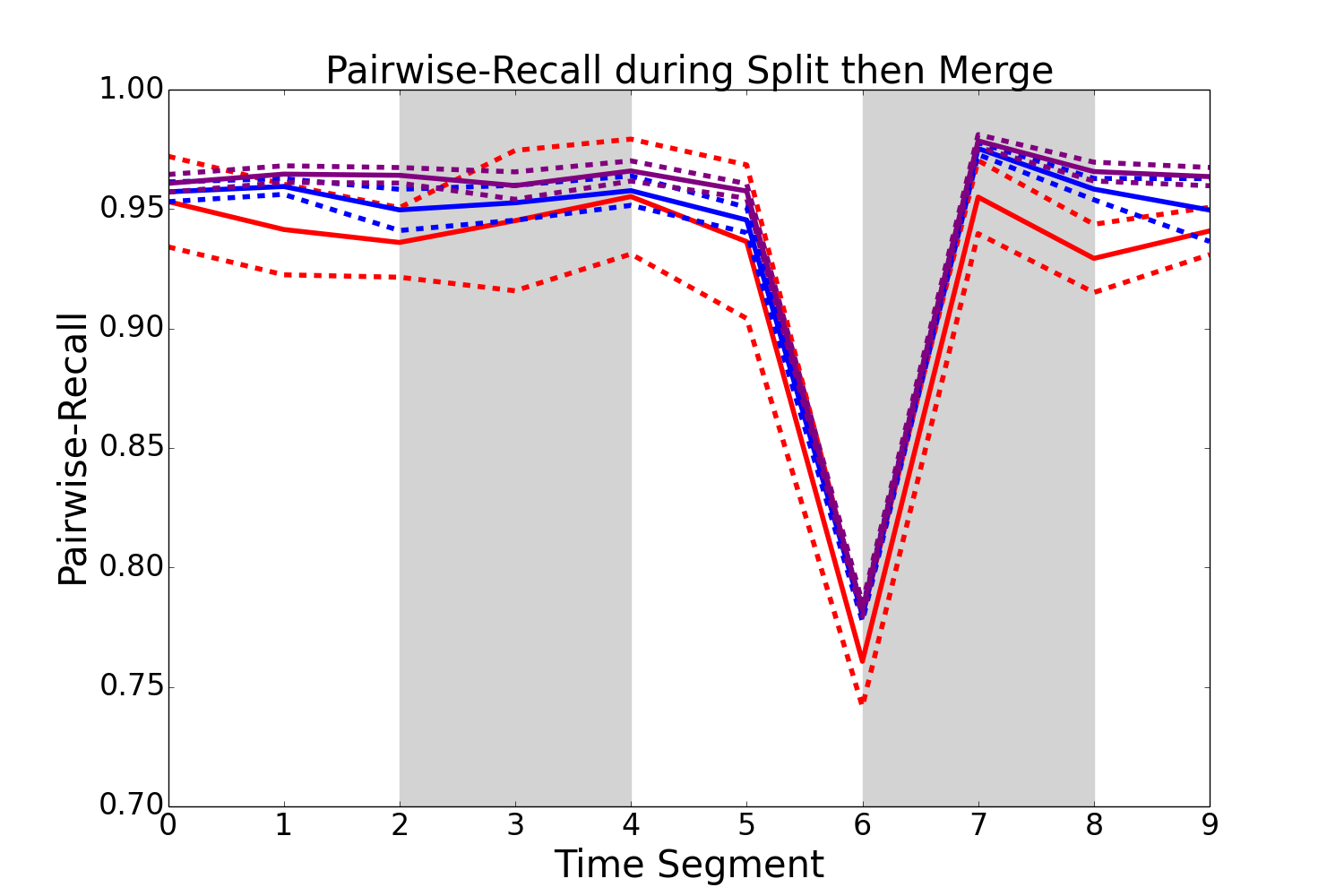}
\end{subfigure}
\caption{\small Performance on synthetic split and merge dynamic event.  Red and blue lines represent the baseline algorithm and the ensemble, respectively, with dashed lines showing the standard error over 10 experiments, as in Figure \ref{experimentemerging}. Purple lines, solid and dashed, show the same for the ensemble approach with smoothing between time segments.  The grey regions represent time periods during which the underlying stochastic block model is changing.  The left-hand panel shows the number of blocks identified in each time segment using the baseline and the ensemble as compared to the ground truth, shown by the black line.  The middle panel shows the pairwise-precision for all three methods over time and the right-hand panel shows the pairwise-recall.}
\label{experimentsynthetic}
\end{figure*}

The results of this experiment are shown in Figure \ref{experimentsynthetic}  Here, the red and blue lines represent the baseline and ensemble algorithms, respectively, as in Figure \ref{experimentemerging}.  The purple lines represent the ensemble with smoothing. The gray regions represent times when the block structure is in flux; the ground truth is set to be that of the blocks at the end of the transition, hence the drop in performance at the start of each region when the behavior observed has yet to fully align with the new ground truth. 

The left-hand panel in Figure \ref{experimentsynthetic} reports the number of clusters found by the various algorithms used.  The black line represents the number of clusters in the ground-truth clustering.  The middle panel reports the pairwise-precision over time and the right-hand panel shows the pairwise-recall.  

At the start of the split event, the precision drops as the ground truth has more clusters than the immediate graph structure (until the cluster has fully split).  As the clustering algorithm will misidentify nodes in the splitting cluster as being in the same community, the precision suffers.  Likewise at the start of the merge event the number of communities is underestimated, placing members of the merging communities as separate which causes recall to drop.

For number of blocks found, the ensemble method performs slightly better than the baseline, especially in more quickly identifying a split.  The smoothed ensemble matches the ground truth cluster number almost exactly.  In terms of the pairwise-precision, the ensemble algorithm has strictly superior precision to the basic algorithm while the smoothed algorithm is better still.  The same relationship holds for recall but the difference is not statistically significant at all times.

\subsection{Semi-Synthetic Experiment}
\label{sec:semireal_experiments}

\begin{figure*}
%\begin{center}
\begin{subfigure}[b]{0.33\textwidth} 
\includegraphics[width=\textwidth]{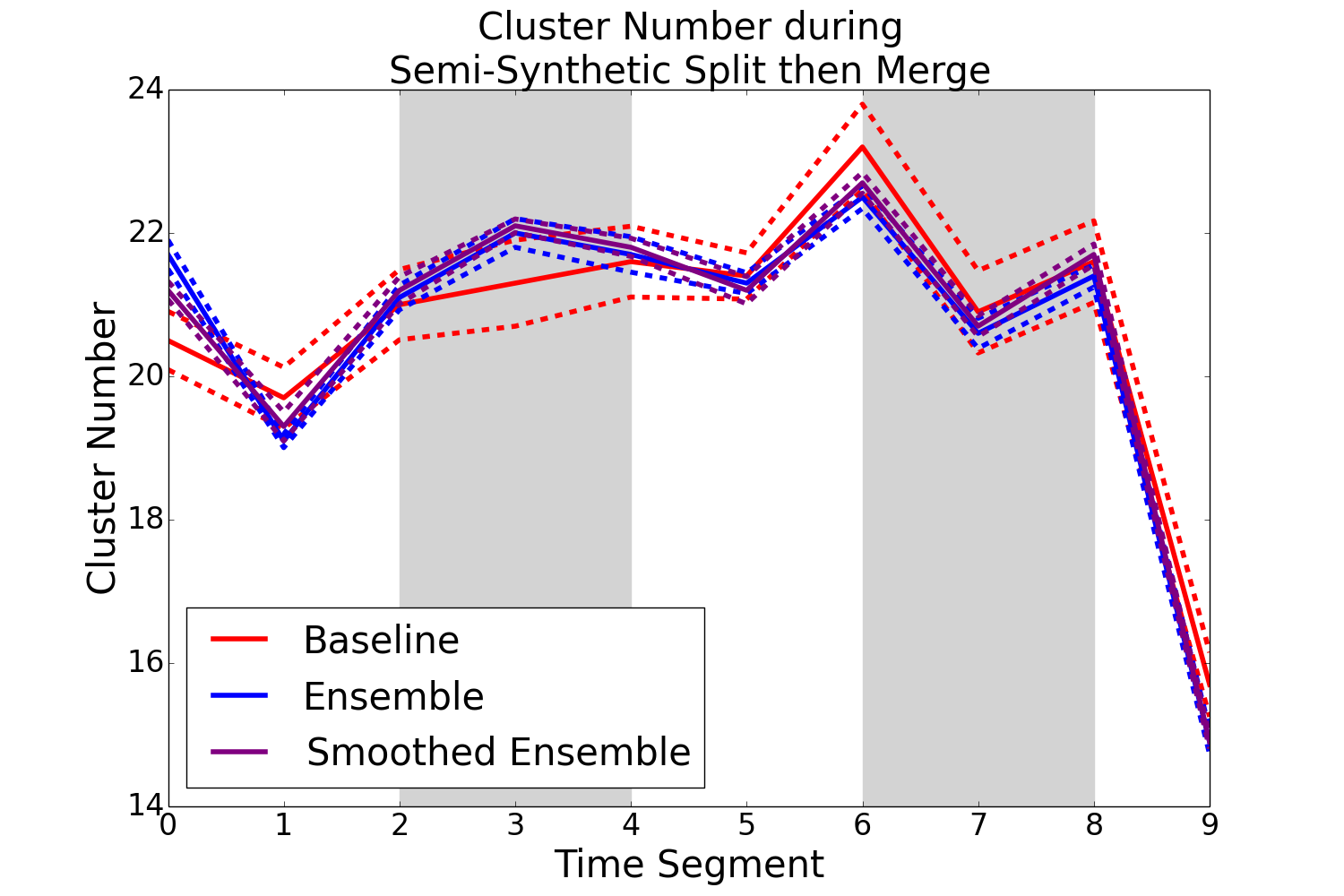}
\end{subfigure}
\begin{subfigure}[b]{0.33\textwidth} 
\includegraphics[width=\textwidth]{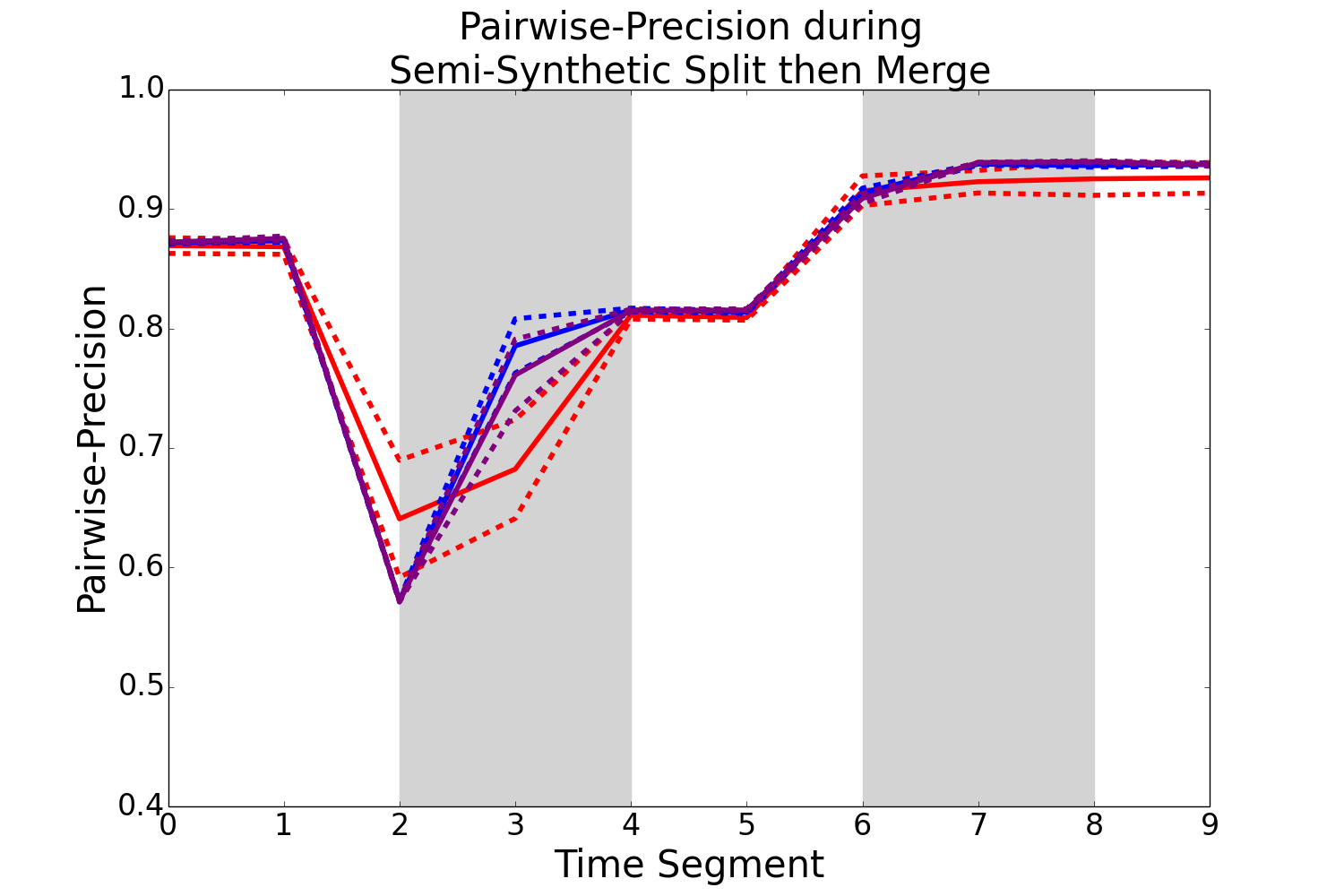}
\end{subfigure}
\begin{subfigure}[b]{0.33\textwidth} 
\includegraphics[width=\textwidth]{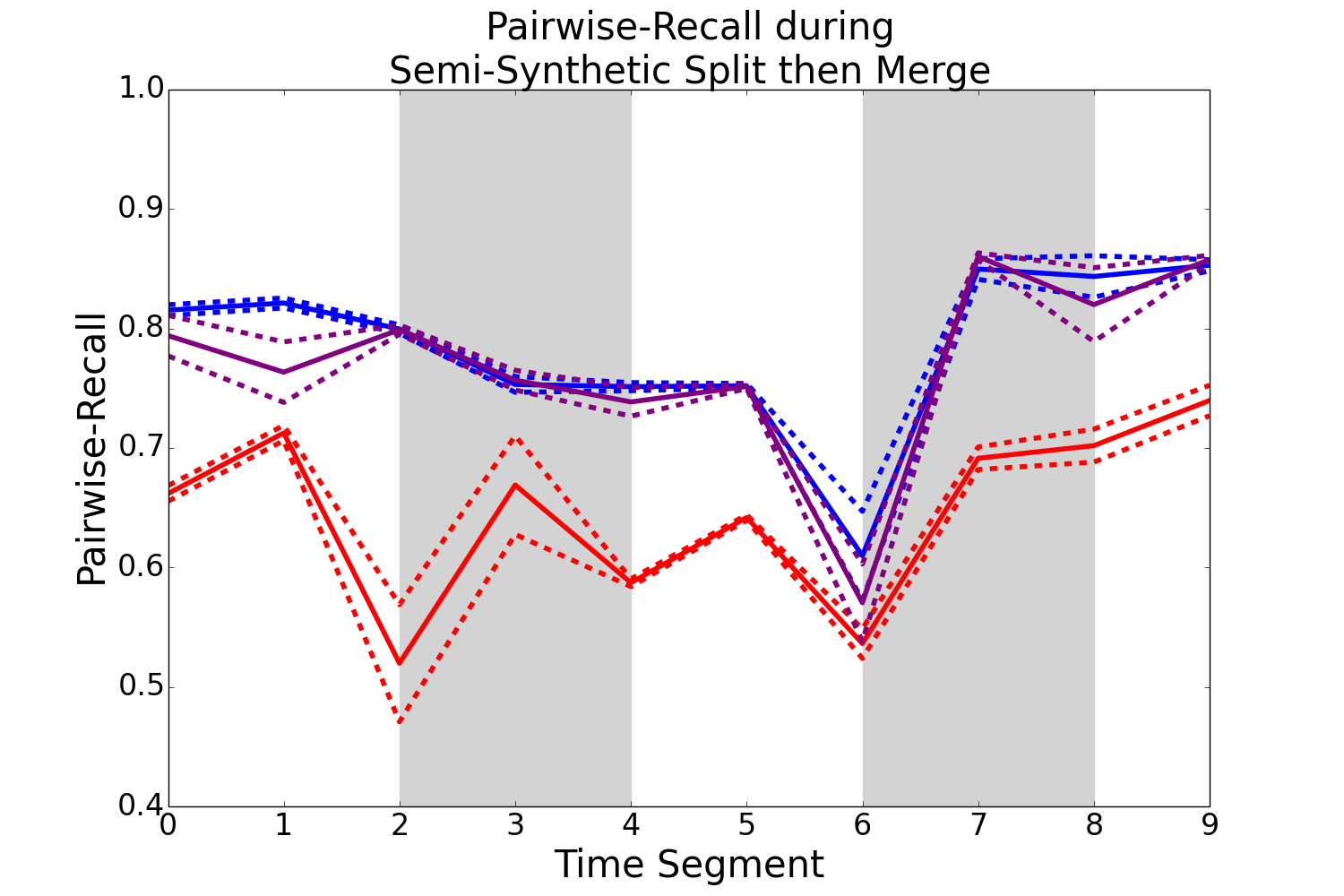}
\end{subfigure}
\caption{\small Performance on semi-synthetic split and merge dynamic event. Red, blue, and purple lines represent the baseline MCMC clustering algorithm, the ensemble approach, and the ensemble approach with smoothing, respectively, with the dashed lines showing the standard error over 10 runs.  The grey regions represent time periods during which the underlying stochastic block model on the synthetic nodes is evolving.  The left-hand panel shows the number of clusters found by each algorithm over time, the middle panel shows the pairwise-precision for the synthetic nodes, and the right-hand panel shows the pairwise-recall for the synthetic nodes.}
\label{experimentsemisynthetic}
%\end{center}
\end{figure*}

The third experiment is performed on a semi-synthetic dataset which was produced by inserting nodes and edges sampled from a block model into a real-world network, the ``email-Eu-core-temporal" dataset from the SNAP database \cite{SNAP}.
The synthetic blocks inserted consist of a 100-node block which splits into two 50-node blocks between time segments 2 and 4 and two 50-node blocks which merge into one 100-node block between time segments 6 and 8.
This is similar to the dynamic structure in the previous experiment, except that now nodes from the synthetic blocks are randomly connected to nodes in the real-world email dataset, rather than nodes from other synthetic blocks.  This more accurately measures the ability of the algorithms to identify dynamic community structure under conditions which more closely represent those found in real-world data.

The synthetic data has 960 edges per time segment, 800 of those being between pairs of synthetic nodes and 160 going to nodes in the real-world data; the real-world data consists of the first 10 weeks of the dataset which has 986 nodes and 38330 edges, split fairly evenly across those weeks.  Pairwise-precision and pairwise-recall are reported for the synthetic nodes only.

The results of this experiment on semi-synthetic data are shown in Figure \ref{experimentsemisynthetic}.  As previously, the red, blue, and purple lines represent the baseline MCMC, ensemble, and ensemble with smoothing algorithms, respectively.  The left-hand panel shows the number of clusters reported by all three algorithms.  Clusters discovered in the real data are included in this count, which is the cause of the drop in cluster number in the final time segment: two large merges occurred in the real dataset at this time reducing the overall number of clusters.  These merges are visible in the community structure shown in Figure \ref{experimentEU}.

The grey bars highlight the time segments in which the clusters of synthetic nodes are undergoing a split or merge.  Unlike the previous experiment, the ground-truth number of clusters is not reported due to the fact that the ground truth communities are not known \emph{ a priori} for the  ``email-Eu-core-temporal" dataset.  The middle and right-hand panels show the pairwise-precision and pairwise-recall over time.
In this experiment the pairwise-precision does not seem significantly different across the different algorithms, however the recall for the baseline algorithm is much worse than either the ensemble or the smoothed ensemble.
%\ctxt{Like the previous experiment the basic algorithm has inferior precision; however, the smoothed algorithm does not seem significantly better than the ensemble without smoothing.  While the basic algorithm is inferior at some times in terms of recall, for most time steps the performance of all three methods is not significantly different under this accuracy measure.  Also notable is the lack of a precision or recall dip when the structure undergoes change.  This is likely due to the noise from the real data obscuring the signal created by the transitions in the synthetic structure.}

\subsection{Real-World Experiment}
\label{sec:real_experiments}

%The final experiment, shown in Figure \ref{experimentenron}, is a short snapshot from the Enron e-mail data.  Here we demonstrate the ability to construct a visualization of the community structure from the smoothed ensemble algorithm's results.  This snapshot shows the period from June 1999 to June 2000, during which the first major split in the company's community structure occurs.

\begin{figure}
\begin{center}
%\begin{subfigure}[b]{0.5\textwidth} 
%\includegraphics[width=\textwidth]{"cluster-number-synth".png}
%\end{subfigure}
\begin{subfigure}[b]{0.475\textwidth} 
%\fbox{\includegraphics[width=\textwidth,trim={2cm 17cm 14cm 10cm},clip=true]{"enron_modified_deleted".pdf}}
%\fbox{\includegraphics[width=\textwidth,trim={0cm 0cm 0cm 0cm},clip=true]{"email-Eu-timeline-short-modified".pdf}}
\fbox{\includegraphics[width=\textwidth,trim={0cm 0cm 0cm 0cm},clip=true]{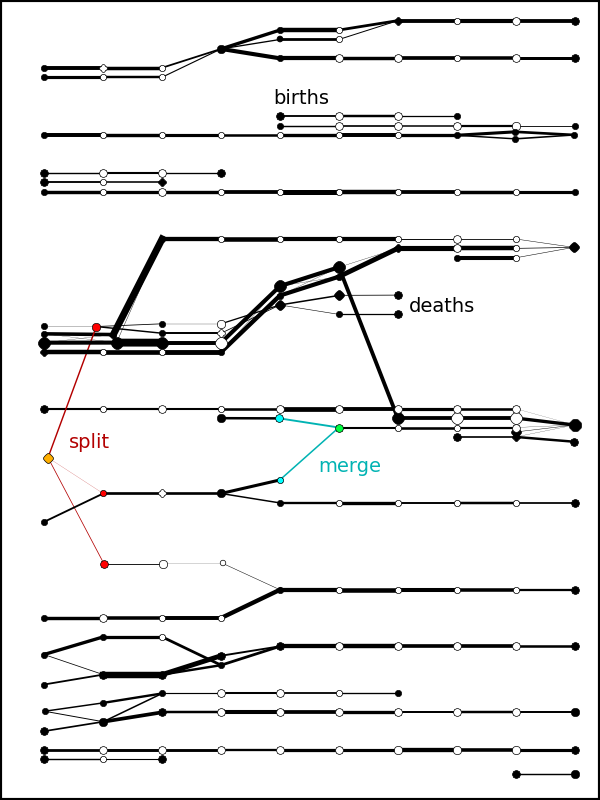}}
\end{subfigure}
%\begin{subfigure}[b]{0.35\textwidth} 
%\includegraphics[width=\textwidth]{"pairwise-recall-semisynth".png}
%\end{subfigure}
\caption{\small Timeline of the clustering behavior of SNAP European University dataset over the first 10 weeks of data using the ensemble algorithm with smoothing.}
\label{experimentEU}
\end{center}
\end{figure}

Finally, we ran the smoothed ensemble framework on the first 10 weeks of the EU email dataset.  The dynamic community structure obtained is shown in Figure \ref{experimentEU}.  The horizontal axis represents time (in weeks), with community detection taking place weekly. The points represent the communities found during each week, with the diameter of the points representing the size of the communities at a given time and the thickness of the lines between communities in different weeks representing the node set overlap between two communities.  Many of the communities are constant, forming long horizontal lines with no splitting or merging behavior.  Others undergo prominent splitting and/or merging behavior.  There are also a number of communities which appear to ``die'' over the course of the ten weeks, with the community terminating without merging into another.  Of particular note are the pair of large merge events occurring in the final time segment where multiple communities join to form large blocs.

\section{Conclusions}

The ability to perform dynamic clustering on real-world networks and produce an interpretable visualization is a powerful tool for network analysis.  The ensemble framework presented in this paper has the ability to produce high-level summations of the evolving community structure of dynamic networks and can be applied to large networks due to its efficient parallel implementation .  We also presented a version of the ensemble with memory which allows for more stable representations of the community structure.  We demonstrated using synthetic, semi-synthetic and real-world scenarios that the ensemble algorithm has superior precision and recall performance compared to the basic clustering algorithm, and we also presented the types of dynamic community representations that the ensemble algorithm can produce from real graphs.

Future directions for this framework include incorporating more information from previous time segments when performing clustering: for example, edges from graphs in previous segments could be included in the current graph using an exponential decay scheme.  This would allow a more narrow resolution of the time segments without making the graphs so sparse as to be useless.  Other possible improvements exist for the visualization of the dynamic network: for example, the variability in the node membership of the representative clusters could be visualized to represent the level of ``confidence" the algorithm has in a particular cluster assignment.

\vspace{2mm}

\noindent \textbf{Acknowledgements}

\vspace{2mm}

This work was performed under the auspices of the U.S. Department of Energy by Lawrence Livermore National Laboratory under Contract DE-AC52-07NA27344.

%\ctxt{
%Ensemble approach useful for stochastic clustering algorithms
%better performance 
%Gives good visualization
%Future work: more incorporation of past graph data other than smoothing ; 
%improvements to visualization and post-processing output 
%TODO: write conclusions section
%}

%bibliography

\bibliographystyle{ieeetr}  
 \bibliography{GraphChallenge}

\end{document}